\documentclass[showpacs,preprintnumbers,amsmath,amssymb,prl,superscriptaddress,twocolumn]{revtex4-1}

\usepackage{txfonts}
\usepackage{graphicx}
\usepackage{dcolumn}
\usepackage{bm}
\usepackage{color}
\usepackage{subfigure}  
\def\comment#1{}

\def\slashchar#1{\setbox0=\hbox{$#1$}           
	\dimen0=\wd0                                 
	\setbox1=\hbox{/} \dimen1=\wd1               
	\ifdim\dimen0>\dimen1                        
	\rlap{\hbox to \dimen0{\hfil/\hfil}}      
	#1                                        
	\else                                        
	\rlap{\hbox to \dimen1{\hfil$#1$\hfil}}   
	/                                         
	\fi}                                         %

\def\nablab{{\mbox{\boldmath $\nabla$}}}

\newcommand{\red}{\textcolor{black}}

\begin{document}

\title{Josephson currents induced by the Witten effect}
\author{Flavio S. Nogueira}
\affiliation{Institute for Theoretical Solid State Physics, IFW Dresden, Helmholtzstr. 20, 01069 Dresden, Germany}
\affiliation{Institut f{\"u}r Theoretische Physik III, Ruhr-Universit\"at Bochum,
Universit\"atsstra\ss e 150, 44801 Bochum, Germany}

\author{Zohar Nussinov}
\affiliation{Physics Department, CB 1105, 
Washington University, 1 Brookings Drive, St. Louis, MO 63130-4899}

\author{Jeroen van den Brink}
\affiliation{Institute for Theoretical Solid State Physics, IFW Dresden, Helmholtzstr. 20, 01069 Dresden, Germany}
\affiliation{Department of Physics, Harvard University, Cambridge, Massachusetts 02138, USA}

\date{Received \today}

\begin{abstract}
We reveal the existence of a new type of topological Josephson effect involving type II superconductors and three-dimensional topological  insulators as tunnel junctions.  We predict that vortex lines induce a variant of the Witten effect that is the consequence of the axion electromagnetic response of the topological insulator: at the interface of the junction each flux quantum attains a fractional electrical charge of e/4.
As a consequence, 
if an external magnetic field is applied perpendicular to the junction, the Witten effect induces an AC Josephson effect in absence of any external voltage. We derive a number of further experimental consequences and propose potential setups where these {quantized, flux induced, Witten} effects may be observed.   
\end{abstract}

\maketitle

One distinguished feature of topological superconductivity is that it hosts gapless boundary states in the form of Majorana fermions, i.e., particle 
states that are their own antiparticles \cite{Wilczek-2009}. One way of indirectly detecting Majorana fermions, and therefore the topological nature 
of superconductivity in the system, was proposed by Kitaev long time ago in 
the framework of a simple, exactly solvable, model \cite{Kitaev-2001}. Kitaev pointed out that 
a semiconducting wire with strong spin-orbit coupling, where $p-$wave-like superconductivity is induced by proximity-effect 
to an $s-$wave superconductor, could host unpaired Majorana modes at its ends, provided the chemical 
potential does not exceed the energy gap between the elementary excitations. Within this framework, a tunnel junction between 
two Kitaev wires would feature fused Majorana modes and lead to a fractional ($4\pi$-periodic) Josephson effect. Since Kitaev's seminal paper, 
several papers have discussed further effects and the possible use of Majorana fermions as the means to realized topologically 
protected quantum computation \cite{Alicea}. Experimental evidence for a fractional AC Josephson effect has been reported in a hybrid 
InSb/Nb nanowire \cite{InSb/Nb nanowire}, thus providing evidence for fused Majorana states.  

Another possibility to realize a fractional Josephson effect is having a topological insulator (TI) as a tunnel junction  \cite{Fu-Kane-2009}. 
In this case, a fractional Josephson effect due to fused Majorana fermions also emerges as a consequence of the proximity effect. Recently, 
the AC Josephson effect has also been measured in this case using HgTe as the three-dimensional insulator junction \cite{4pi-Josephson-HgTe}. 
Despite the recent progress in the measurement of Josephson effect phenomena related to Majorana physics, the true topological character has 
yet to be clearly demonstrated. For instance, 
it remains to be shown that non-Abelian statistics can be realized in some way by means of Josephson junctions featuring 
fused Majorana modes, which would pave the way to implement quantum information processing \cite{Alicea-2011}.  

In this Letter, we show that another type of topological Josephson effect is also present in SC-TI-SC junctions (SC=superconductor), when 
an external magnetic field is applied {\it perpendicular} to the junction and the superconductor is a type II one. We will show 
that the induced vortex lines act as magnetic monopoles in the sense that they trigger a variant of the Witten effect, which in a field theory setting endows magnetic monopoles with a fractional electric charge \cite{Witten}. In our condensed matter setting the resulting electrical charge of magnetic vortices is the consequence of the axion electromagnetic response of the topological insulator \cite{Qi-2008}.
This has a number of experimentally accessible consequences. As the vortex lines perpendicular  
to the TI junction become electrically polarized they may trigger an AC Josephson effect in {\it absence} of an external voltage. 
Furthermore, the Josephson frequency will turn out to be quantized, as a consequence of a Berry phase for the tunnel junction 
induced by the Witten effect. This will, in turn, imply a peculiar behavior for Shapiro steps. These Josephson-Witten effects are expected to be rather robust as they merely rely on magnetic fluxes traversing the SC-TI boundary and do not require any further fine-tuning or interplay between different order parameters.

{\it ---Electromagnetic variant of the Witten effect} 
For a strong three-dimensional TI, it was shown \cite{Qi-2008} that the electromagnetic response leads to a so called axion term \cite{Wilczek}
in the effective electromagnetic Lagrangian, 
\begin{equation}
{\cal L}_{\rm Axion}=\frac{e^2\theta}{4\pi^2}{\bf E}\cdot{\bf B},
\end{equation}
where units are such that $\hbar=c=1$. Generally, $\theta$ is a field, the so called axion. 
Several properties are important in the following~\cite{Sitte2012}: (i) The value of  $\theta$ is defined modulo $2\pi$; (ii) within the bulk $\theta$ is constant; (iii) For a uniform $\theta$, the above Lagrangian  is a total derivative, yielding henceforth a surface term in the action; and (iv) in the presence of time-reversal symmetry (TRS),  $\theta$ can either be $0$ (the topologically trivial case) or $\pi$: in particular for a TRS strong topological insulator $\theta=\pi$. However in the absence TRS, $\theta$ need not be quantized, and a TI can adiabatically be transformed into a band insulator.
Since the axion term is a surface term, it does not changes 
the field equations. However, it does change the boundary conditions. This point is crucial and leads to the Witten 
effect \cite{Witten}: when magnetic monopoles are present the total electric charge becomes fractionalized due to the magnetic monopoles 
acquiring  an electrical charge. 

The Witten effect has been originally predicted for an $O(3)$ Higgs model, which has magnetic monopole solutions 
in the spontaneous symmetry breaking regime as 
demonstrated by 't Hooft and Polyakov \cite{tHooft-Polyakov}. However, in electrodynamics, magnetic monopoles have to be added 
by hand. Nevertheless, in the presence of the axion term, the Witten effect also holds \cite{Wilczek} just as in the more general case originally 
discussed by Witten. Furthermore, as we will show in the following, a form of the Witten effect also holds for the case of vortex lines.  
To this end, let us consider the simplest action for electrodynamics with an axion term,
\begin{equation}
	\label{Eq:axion-ED}
	S=\int dt\int d^3x\left[\frac{1}{8\pi}({\bf E}^2-{\bf B}^2)+\frac{e^2\theta}{4\pi^2}{\bf E}\cdot{\bf B}-\rho\phi-{\bf j}\cdot{\bf A}\right].
	\end{equation}
In a system without boundaries, the Witten effect readily follows from the integral form of the Gauss law 
in the presence of the $\theta$-term, 
\begin{equation}
\label{GaussQ}
4\pi Q=\oint_Sd{\bf S}\cdot{\bf E}=4\pi\left(q-\frac{e^2\theta}{4\pi^2}\int_V d^3r\nablab\cdot{\bf B}\right),
\end{equation}
where
where $q$ is the electrical charge, $Q$ the total charge and
the volume integral is bounded by the surface $S$. In standard electrodynamics $\nablab\cdot{\bf B}=0$, such that for constant 
$\theta$ nothing happens, leading to $Q=q$.  
If we assume that the theory has 
magnetic monopoles the standard Witten effect \cite{Witten} arises. Using that the electrical charge $q=ne$ ($n\in\mathbb{Z}$) and the flux of a single monopole $\Phi_B=2\pi/e$ the total charge is
\begin{equation}
Q=e\left(n-\frac{\theta}{2\pi}\right).
\end{equation}

\begin{figure}
	\centering
	\includegraphics[width=1.05\columnwidth]{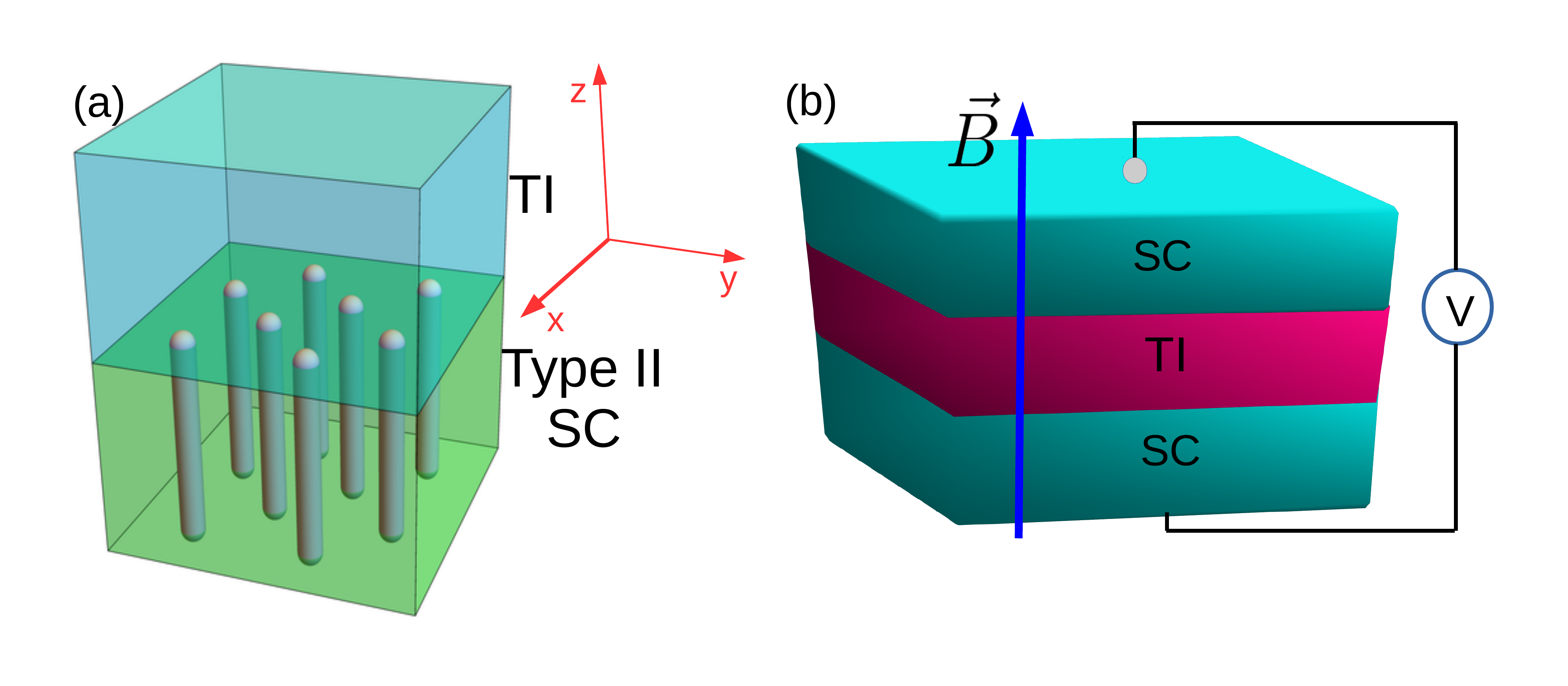}
	\caption{(Color online) {(a) Schematic view of a junction between a strong TI (e.g. Bi$_2$Se$_3$, Bi$_2$Te$_3$ or strained HgTe) and a type II SC (e.g. Nb,V, or a high T$_c$ cuparate). The magnetic flux from the vortex lines at the interface induces a charge fractionalization due to the Witten effect. (b) Schematics of possible experimental setup to measure Josephson-Witten effect. The Witten effect acts on the vortex lines rather than on magnetic monopoles, thus creating a potential difference across the junctions.}
	}
	\label{Fig:TI-SC}
\end{figure}

In the electrodynamics of condensed matter systems magnetic monopoles do not arise. However, the presence of a TI surface prompts $\theta$ 
to change, and the the Gauss law in the form 
 $\nablab\cdot{\bf E}=4\pi\rho+(e^2/\pi)\nablab\theta\cdot{\bf B}$ needs  to be used 
in order to  accommodate this change. Let us now specifically consider the situation of
a half-space ($z<0$) occupied with a type II superconductor interfacing with a strong TI on the upper half-space ($z>0$); see Fig. \ref{Fig:TI-SC}-(a). An external field ${\bf H}_{\rm ext}$ 
perpendicular to the interface $z=0$ generates magnetic vortices in the superconductor. 
Inside the TI and near the interface there are stray fields originating from the vortex lines with the boundary condition 
 that ${\bf B}={\bf H}_{\rm ext}$ for $z\to\infty$. We have that $\theta(z)=0$ for $z<0$ (i.e., inside the superconductor), while 
 $\theta$ is uniform in the TI, including its surface at $z=0$.
For straight vortex lines, the magnetic field inside the superconductor 
depends only on the in-plane radial coordinate, and we find for the charge at the
the $z=0$ interface, 
\begin{eqnarray}
\label{Eq:Qv}
Q&=&q+\frac{e^2}{4\pi^2}\int d^2r B(r)\int_{-\infty}^{0}dz\frac{d\theta}{dz}
\nonumber\\
&=&q+\frac{e^2\theta}{4\pi^2}\Phi_B,
\end{eqnarray}
where now $\Phi_B= N_v \Phi_{0} =N_v\pi/e$ is the total flux due to $N_v$ flux lines. Here,
$\Phi_{0} = 2 \pi/e^* = \pi/e$ is the elementary flux quantum associated with the Cooper pair ($e^*=2e$).
With $d$ the thickness of the TI, a charge $q-\frac{e^2\theta}{4\pi^2}\Phi_B$
is similarly found at the (top) $z=d$ SC-TI interface. This analysis implies that when a TI shares an interface with a type II superconductor, the flux of the vortex lines becomes electrically polarized by the Witten effect.

At the interface of a SC  and a strong TI 
with time-reversal symmetry $\theta=\pi$
the Witten effect endows each magnetic flux with a charge $e/4$. 
\red{Since a vortex is a solitonic object, it is allowed to carry  a fractional electrical charge, similarly to the situation found  
with magnetic monopoles (also solitonic objects) in the standard Witten effect.} It is interesting that the statistical properties of these fractionally charged fluxes might in principle be determined by shot noise \cite{Fradkin-2005} or interferometry \cite{Bonderson-2008} experiments. 

\red{Microscopically, the origin of this charge fractionalization can be traced back in part to the origin of the axion term in 
	the Lagrangian of a three-dimensional TI \cite{Qi-2008}. In this case a well-known argument shows that the electromagnetic 
	response implies a half-quantized Hall conductivity $\sigma_{xy}$ for the surface electrons \cite{Qi-2008}.  Thus, we can adapt a simple argument 
	by McGreevy \cite{McGreevy-lectures} estimating the amount of charge $\Delta Q$  
	acquired by a localized flux to our case as follows. Applying the Faraday law to the 
	elementary flux quantum $\Phi_0=\pi/e$ of a vortex  and noting that there are circulating Hall currents at the vortex core, we obtain, 
	}
\begin{equation}
\Phi_0=\frac{\pi}{e}=-\int dt\oint d{\bf r}\cdot{\bf E}=-\frac{\Delta Q}{\sigma_{xy}},
\end{equation}
\red{where we have used that the transverse current in the vortex core $j_r=\sigma_{xy} E_\varphi$. Since 
	$\sigma_{xy}=-e^2/(4\pi)$ (this is the value of the half-quantized 
	Hall conductivity, $\sigma_{xy}=e^2/(2h)$ when $\hbar=1$), the above simple argument yields $\Delta Q=e/4$.}

\red{The discussion above can be further elaborated to enable to consider 
 time-dependent magnetic fluxes, yielding  
 another instance of the Witten effect}. In this case the action (\ref{Eq:axion-ED}) becomes 
\begin{eqnarray}
S&=&\int dt\int d^3x\left[\frac{1}{8\pi}({\bf E}^2-{\bf B}^2)-\rho\phi-{\bf J}\cdot{\bf A}
\right.\nonumber\\
&-&\left.\frac{e^2}{4\pi^2}\nablab\cdot(\theta{\bf E}\times{\bf A})\right],
\end{eqnarray}
where the total current density, 
\begin{equation}
{\bf J}= {\bf j}-\frac{e^2}{4\pi^2}(\nablab\theta\times{\bf E})-\frac{e^2\theta}{4\pi^2}\nablab\times{\bf E}.
\end{equation}
In the above equation the term $\sim (\nablab\theta\times{\bf E})$ contributes to the Hall conductivity while the term 
$\sim \theta\nablab\times{\bf E}$ contributes to the Witten effect. For example, if ${\bf E}$ is uniform and applied in the $x$-direction parallel to the TI   
surface, we obtain the Hall current along the $y$-direction, 
\begin{equation}
j_y^H=\frac{e^2E}{4\pi^2}\int_{0}^{d}dz\frac{d\theta}{dz}=\frac{e^2E}{4\pi^2}[\theta(d)-\theta(0)],
\end{equation}
where $d$ is the thickness of the TI. 
On the other hand, if an electric field is induced due to an external magnetic field applied perpendicular to the surface, 
the total electric current flowing through one of the surfaces is given by, 
\begin{eqnarray}
\label{dqdt}
\frac{dQ}{dt}&=&-\int_S d{\bf S}\cdot{\bf j}+\frac{e^2\theta}{4\pi^2}\int_Sd{\bf S}\cdot(\nablab\times{\bf E})
\nonumber\\
&=&\frac{dq}{dt}+\frac{e^2\theta}{4\pi^2}\oint_Cd{\bf r}\cdot{\bf E}
=\frac{dq}{dt}-\frac{e^2\theta}{4\pi^2}\frac{d\Phi_B}{dt},
\end{eqnarray}
where we invoked Faraday's law.  
Eq. (\ref{dqdt}) thus indeed extends our Witten effect results to the time dependent domain. 
This analysis implies that when a TI shares an interface with a type II superconductor, the flux of the vortex lines becomes electrically polarized by the Witten effect.

{\it ---Capacitance due to Witten effect} 
The magnetic flux carrying a fractionalized charge due to the Witten effect highlights the interplay between topology and boundary 
conditions, since the axion term is a total derivative. An immediate consequence of this interplay is the modification of the capacitance 
energy.  One expects that due to the Witten effect a capacitor can become electrically charged by magnetic fluxes in the plates. 
Indeed, an application of the Gauss law to a parallel plate capacitor yields the magnitude of the electric field, 
\begin{equation}
E=\frac{4\pi}{A\epsilon}\left(q-\frac{e^2\theta}{4\pi^2}q_m\right),
\end{equation}       
where $\epsilon$ is the dielectric constant of the material filling the capacitor, 
$q_m$ is the total magnetic monopole charge, and $A$ is the area of the plate. 
Thus, irrespective of the plate separation, we obtain the voltage difference, 
\begin{equation}
\label{Eq:V-monopole}
\Delta V=\frac{1}{C}\left(q-\frac{e^2\theta}{4\pi^2}q_m\right),
\end{equation}
where $C$ is the usual capacitance of a parallel plate capacitor.  We see that for non-zero $q_m$ the voltage difference is nonzero even if there are no 
electric charges in the capacitor.
One important feature of this capacitor is that the monopole charges in one plate are connected by vortex strings to the opposite monopole 
charges on the other plate. The Witten effect implies that this vortex string is electrically polarized. 

{\it ---Josephson-Witten effect} 
Having established that magnetic monopoles and fluxes can generate an electrical potential difference in presence of a varying axion field, we consider a SC-TI-SC Josephson junction in this context.  Magnetic monopoles are of course absent in this physical situation. The Hamiltonian of a Josephson junction \cite{Tinkham-Book} features 
the charging or capacitive energy, $C(\Delta V)^2/2$ and the Josephson 
potential energy, $-E_J\cos\Delta\phi$. The voltage difference can be related to the variation of particle number, $\Delta n$, which is 
semi-classically conjugate to the phase difference, $\Delta\phi$, via, $\Delta V=(2e/C)\Delta n$.  Thus \cite{Tinkham-Book}, 
\begin{equation}
H_J= \frac{2e^2}{C}(\Delta n)^2-E_J\cos\Delta\phi.
\end{equation}
For a SC-TI-SC Josephson junction, an additional surface energy has to be included due to the 
axion term, and the Hamiltonian of the junction modifies to,  
\begin{equation}
\label{Eq:HJ}
H_J= \frac{2e^2}{C}\left(\Delta n+\frac{e\theta}{8\pi^2}\Phi_B\right)^2-E_J\cos\Delta\phi,
\end{equation}
since due to the 
$\theta$-term a charging energy is induced  
when an external magnetic field is applied {\it perpendicular} to the junction. As usual, 
$\Delta n$ is the Cooper pair number variation that is conjugate to the phase difference $\Delta\phi$ across the junction.
Therefore, $i [H_J,\Delta \phi]= \partial_t\Delta\phi=2e\Delta V_{\rm ind}$ with the induced potential drop 
\begin{equation}
\label{Eq:dVind}
\Delta V_{\rm ind}=\frac{1}{C}\left(2e\Delta n+\frac{e^2\theta}{4\pi^2}\Phi_B\right).
\end{equation}
Since $\Phi_B$ is quantized in multiples of $\Phi_{0}=\pi/e$, we see from Eqs.(\ref{Eq:HJ}) and (\ref{Eq:dVind}) that the original $2\pi$-periodicity of $\theta$, which is intrinsic to TI electrodynamics \cite{Qi-2008}, 
becomes a $8\pi$-periodicity, upon proximity with the superconductor. 
Indeed, we obtain that for arbitrary $N_v$ that both 
Eqs. (\ref{Eq:HJ}) and (\ref{Eq:dVind}) are invariant under $\theta\to\theta+8\pi$, $\Delta n\to\Delta n-N_v$. This periodicity is also 
reflected in the excitation spectrum of the Hamiltonian (\ref{Eq:HJ}) in the case where the charging energy dominates over the Josephson 
energy $E_J$. In this case an elementary second-order perturbation theory calculation yields, 
\begin{equation}
\label{Eq:spectrum}
E_n(\theta)\approx\frac{2e^2}{C}\left(n+\frac{\theta}{8\pi}N_v\right)^2+\frac{(E_JC)^2/(2e^4)}{4\left(n+\frac{\theta}{8\pi}N_v\right)^2-1},
\end{equation}
where $n\in\mathbb{Z}$. For $\theta=0$ Eq. (\ref{Eq:spectrum}) just reduces to the perturbation solution of the two-dimensional Stark rotator. 
{For $\theta\neq 0$ the problem is equivalent to one of a particle moving on a ring with a magnetic flux 
	$\Phi=\theta N_v/4$, with tunnel barrier \cite{Bloch}. {This equivalence implies that for $N_v=1$ an electron going around the ring with}  a single vortex will {effectively only pick up } a flux $\theta/4$, which in the case of a time-reversal invariant system means a $\pi/4$ flux.} 
From this 
point of view,  gauge invariance and the $8\pi$ periodicity of $\theta$ imply the invariance 
of the spectrum (\ref{Eq:spectrum}) under the transformation, 
$n\to n+m$, $\theta\to\theta-8\pi m/N_v$.  This result holds exactly, being independent of the perturbative calculation, as it is 
a consequence of the gauge invariance of the system.

Let us estimate the Witten effect contribution to the  voltage drop from Eq. (\ref{Eq:dVind}). Candidate materials are structures 
	involving either Bi$_2$Se$_3$-Vanadium or Bi$_2$Te$_3$-Niobium interfaces. In the absence of external voltages, the voltage drop is 
	due uniquely to the Witten effect induced by the vortices, yielding $\Delta V_{\rm ind}=N_ve/(4C)$. 
	Generally, $C$ represents the combined capacitance between TI and superconductor. 
For a small junction the capacitance is in the order of $\sim 1$ pF  \cite{Pablo}. A magnetic field of $\sim 10$ T  gives rise to vortices with a typical spacing of 100 nm, which then induce voltage drop $\sim 2~ \mu{\rm V}$, a drops that is comparable to those found in larger junctions and which is easily within the experimental  limit of detection.

The AC Josephson effect oscillation frequency exhibits an additional contribution due to the $\theta$-term. 
Thus, the total frequency for the AC Josephson effect is $\omega=\omega_0+\omega_\theta$, where $\omega_0=2eV_0$ is the 
usual Josephson frequency with $V_0=2e\Delta n/C$, and, 
\begin{equation}
\label{Eq:Witten-f}
\omega_\theta=\frac{e^3\theta}{2\pi^2C}\Phi_B,
\end{equation} 
is the contribution to the frequency which is induced by the Witten effect. Thus, similarly to Shapiro steps \cite{Shapiro}, we find a DC Josephson 
effect by tuning the voltage such that $\omega_0=-\omega_\theta$. Using our estimate for the voltage drop at high magnetic fields, we 
obtain $\omega_\theta\sim 1$ GHz. Thus, 
in the absence of external voltage, a topological  magnetoelectric contribution is still present, 
implying the AC Josephson current, 
\begin{equation}
I_J(\Delta\phi,t)=2eE_J\sin\left(\Delta\phi+\omega_\theta t\right). 
\end{equation}

The Witten effect has further important consequences if we consider the Josephson effect in small junctions.
To see this let us consider the partition function of the Lagrangian $L_J=\Delta n\partial_t\Delta\phi-H_J$ in terms
of a path integral in imaginary time,
\begin{equation}
\label{Eq:Z}
Z=\int{\cal D}\Delta n{\cal D}\Delta \phi e^{-\int_0^\beta d\tau(i\Delta n\partial_\tau\Delta\phi+H_J)},
\end{equation}
where we have used the fact that $\Delta n$ is canonically conjugate to $\Delta\phi$. Due to the periodicity of $\Delta\phi$, the 
above path integral is calculated with a periodic boundary condition taking the form, $\Delta\phi(\beta)-\Delta\phi(0)=2\pi n_W$, 
$n_W\in\mathbb{Z}$ is 
the standard winding number that arises in the partition function of small junctions \cite{Schoen} and which is due to $\Delta\phi$ being conjugate to the  particle number {operator}.
By performing the shift $\Delta n\to \Delta n-e\theta\Phi_B/(8\pi^2)$, Eq. (\ref{Eq:Z}) acquires a phase factor, 
$e^{ie\theta\Phi_B/(8\pi^2)\int_0^\beta d\tau\partial_\tau\Delta\phi}$,  which contains the integral over a total derivative. 
Due to the boundary condition 
\begin{equation}
\theta=\frac{8\pi m}{N_v}, ~~~~~m\in\mathbb{Z},
\end{equation}
and the $8\pi$ periodicity of $\theta$ corresponds to the translation $m\to m+N_v$.  This result implies, 
\begin{equation}
\omega_\theta=\frac{(2e)^2 m}{C},
\end{equation}
leading to a quantized DC component. Thus, due to the Witten effect, 
the voltage is quantized in a way similar to Shapiro steps \cite{Shapiro}. This result modifies in turn the way the actual Shapiro steps behave, 
since now the phenomenon will be characterized by two integers.  
Indeed, by considering an additional AC voltage, $V(t)=V_0+V_1\cos(\omega_1t)$, the 
standard argument for Shapiro steps \cite{Tinkham-Book}  implies the DC voltages, 
\begin{equation}
\label{Eq:Vnm}
V_{nm}=\frac{n\omega_1}{2e}-\frac{2me}{C}, 
\end{equation} 
in which case the usual Shapiro step result is obtained for $m=0$. 
Eq. (\ref{Eq:Vnm}) leads to a charge lattice, 
$Q_{nm}=CV_{nm}$ reminiscent of the Schwinger result \cite{Schwinger-1969} generalizing the Dirac quantization to dyons, namely, 
dipoles involving an electric and a magnetic charge.  Similarly to that case, we can express the charge obtained from the 
voltage (\ref{Eq:Vnm}) in terms of modular transformations \cite{Cardy-theta} describing a so-called S-duality \cite{Witten-S-duality}. 
There is also a similarity between Eqs. (\ref{Eq:Vnm}) and (\ref{Eq:V-monopole}), when the Dirac duality relation, $qq_m=2\pi$ is 
accounted for. 
 
 
A possible experimental setup to test the above prediction on the quantization of Shapiro steps is shown in Fig. \ref{Fig:TI-SC}-(b). The strategy to 
 observe the result (\ref{Eq:Vnm}) is to measure the $I-V$ 
 characteristics using microwave radiation and an external magnetic field perpendicular to the junction. For the case of a 
 resistively shunted junction, 
 {$I=I_{c0}\sin\Delta\phi+R^{-1}(V+2em/n_W)+CdV/dt$, where $I_{c0}=2eE_J$ is the critical current in absence of the axion term. 
 	Thus,  }
 the small capacitance regime  is  
 shown to {obey the differential equation,
 	\begin{equation}
 	\frac{d\Delta\phi}{dt}=2eI_{c0}R\left[\frac{1}{I_{c0}}\left(I-\frac{2me}{RC}\right)-\sin\Delta\phi\right],
 	\end{equation}
 	which yields the time-averaged voltage,}
 \begin{equation}
 V_m=R\sqrt{\left(I-\frac{2me}{RC}\right)^2-I_{c0}^2},
 \end{equation}
 {which implies a quantized critical current.}
  Since the capacitance is small
 we obtain for small $I$ and $I_{c0}$ a 
 quantized voltage, $V_m\approx 2me/C$. 
 
{\it ---Conclusions}
We have shown that new types of Josephson effects can arise in tunnel junctions between type II superconductors and strong topological insulators because  
the axion electromagnetic response of the TI causes a Witten effect that endows magnetic fluxes at the SC-TI interface with an electrical charge.
This charge is $e/4$ per elementary flux quantum.
As vortex lines perpendicular to the SC-TI junction become electrically polarized they in turn generate an AC Josephson effect in absence of an external voltage. 
Furthermore, the Witten effect contributes directly to the Josephson frequency which in turn modifies the way the Shapiro steps behave, a phenomenon that is now characterized by {\it two} integers. 
One expects these Josephson-Witten effects to be rather robust experimentally as they in the end only rely on magnetic fluxes traversing the SC-TI boundary and do not require any further fine-tuning.


 
\begin{acknowledgments}
This work was supported by the DFG through the Collaborative Research Center SFB 1143. ZN acknowledges support by the NSF under grant no. CMMT 1411229. JvdB acknowledges support from the MIT-Harvard Center for Ultracold Atoms. 
\end{acknowledgments}

\end{document}